# Giant amplification in degenerate band edge slow-wave structures interacting with an electron beam

Mohamed A. K Othman[1], Mehdi Veysi[1], Alexander Figotin[2], and Filippo Capolino[1a)]

[1]*Department of Electrical Engineering and Computer Science, University of California, Irvine, CA 92697 USA.*

[2]*Department of Mathematics, University of California, Irvine, CA 92697 USA.*

We propose a new amplification regime based on synchronous operation of four degenerate electromagnetic (EM) modes in a slow-wave structure and the electron beam, referred to as super synchronization. These four EM modes arise in a Fabry-Pérot cavity (FPC) when degenerate band edge (DBE) condition is satisfied. The modes interact constructively with the electron beam resulting in superior amplification. In particular, much larger gains are achieved for smaller beam currents compared to conventional structures based on synchronization with only a single EM mode. We demonstrate giant gain scaling with respect to the length of the slow-wave structure compared to conventional Pierce type single mode traveling wave tube amplifiers. We construct a coupled transmission line (CTL) model for a loaded waveguide slow-wave structure exhibiting a DBE, and investigate the phenomenon of giant gain via super synchronization using the Pierce model generalized to multimode interaction.

## I. INTRODUCTION

High power sources utilize slow-wave structures (SWSs) to synchronize the EM mode phase velocity to the same for the charge wave on the electron beam. The synchronization is critical to the energy transfer from the beam to the radio frequency wave serving as basis for high power generation,[1,2]. Helical slow-wave structures are among the earliest realizations of SWS due to their cost efficiency and wide bandwidth[3–5]. When seeking for best conditions for the synchronization, researchers proposed a number of periodic structures as SWS such as those comprising coupled cavities[6] or corrugated waveguides[7]. Recently, metamaterial based SWSs have been proposed[8–10] by loading a waveguide with subwavelength resonant elements, like ring-bars[11] or split-ring resonators[8], with the aim of enhancing the interaction impedance and gain accordingly[2,12]. Such periodic SWSs usually operate at frequencies far from the edge of their band gap. The underlying physical operation of amplification in SWSs is well described by resorting to equivalent transmission lines (TLs) with the charge wave supported by the electron beam modeled as a charge fluid, according to a model introduced by Pierce and his contemporaries[3,4,13].

Despite the approximations of a SWS as an ideal TL and the hydrodynamic modeling of the electron beam as a flow of charges, Pierce theory[2,4,14] is still a reliable method to predict traveling wave tube (TWT)'s behavior as well as to engineer the performance under small-signal approximation. More elaborate descriptions account for non-linear effects in TWTs

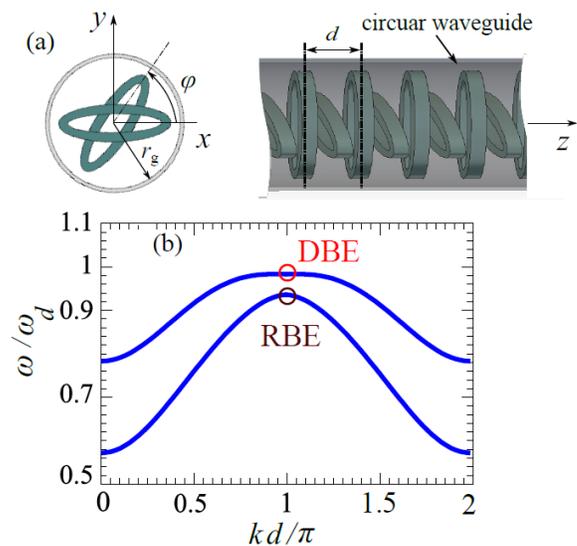

FIG. 1. Example of SWS made of a periodically ring-loaded waveguide that supports a degenerate band edge at the radian frequency $\omega_d$. Though, our TL model is applicable also to other geometries developing DBE. (b) Dispersion diagram obtained by full wave simulation of the "cold" periodic structure in (a), with $k$ being the Bloch wavenumber, and showing two possible band edge conditions, namely the RBE, and the DBE. Note the flatness of the dispersion diagram at $\omega_d$.

[15–17], and a number of numerical methods have been developed incorporating full wave, particle-in-cell (PIC) techniques to accurately model waveguide-beam interaction[18,19]. Typically, periodic SWSs support several Bloch modes (propagating and evanescent), where each mode is represented by an infinite number of spatial Floquet harmonics[20]. Accordingly, if a synchronized spatial harmonic (with a phase velocity matching the beam electron velocity) is amplified, then generically all the

a) f.capolino@uci.edu

other harmonics of that mode would also be amplified.

We propose here a new amplification regime based on synchronous operation of the electron beam and four electromagnetic modes that arise under special degeneracy condition called degenerate band edge (DBE). We refer to this regime as *super synchronization* in contrast to conventional synchronization that involves only a single EM mode. We demonstrate that the super synchronization provides for significantly larger gains for smaller currents and smaller cavity dimensions. The electromagnetic DBE modes have been analyzed in [20,21] and we study here their interaction with an electron beam in finite length devices. A DBE condition as in Fig. 1 causes a quartic power dependence at the band-edge of the dispersion diagram, viz., $(\omega_d - \omega) \propto (k - k_d)^4$, where $\omega_d$ is the DBE angular frequency and $k$ is the Bloch wavenumber. This DBE condition is accompanied by significant reduction in the group velocity of waves and giant improvement in the local density of states [22]. Figotin and Vitebskiy proposed DBE-based frozen mode regimes in a multilayer dielectric one-dimensional lattice [21,23,24], which leads to a dramatic increase in field intensity linked with a transmission band-edge Fabry-Pérot resonance in a finite stack of periodic anisotropic layers with in-plane misalignment. The DBE condition has been realized in microstrip transmission lines reported in [25], and it has been demonstrated recently in fully-metallic circular loaded waveguides [26]. Some theoretical studies have discussed TWT operation very close to the edge of the EM band gap at a regular band edge (RBE), i.e., close to the cut-off frequency [15–17,27–30], which is a different condition than the DBE explored here.

In this premise, super synchronization dictates that the charge wave interacts with all the four modes near and at the DBE that leads to giant amplification (the amplification regime discussed here is also referred to as convective instability scheme [31,32]). We elaborate first on a circular waveguide SWS that is designed to support a DBE, as was demonstrated in [26]. Then we apply a coupled transmission line (CTL) formulation [20,33] in order to explore the enhanced interaction mechanism between the DBE mode and the charge wave. Throughout this paper, we assume the signal to be time-harmonic with a $e^{j\omega t}$ time convention, and all parameters used in numerical simulations are listed in both Appendix A and Appendix B.

## II. CTL MODEL OF A PERIODIC WAVEGUIDE WITH A DBE INTERACTING WITH AN ELECTRON BEAM

A common starting point for engineering an amplifying SWS is to optimally position the dispersion relations of uncoupled "cold" slow wave structure and the electron beam. We have found that realistic dispersion diagram for the coupled/interacting SWS and electron beam ("hot structure") is a significant alteration of the "cold structure" dispersion relations in a critical area. Conventional designs also presume synchronous operation of a single electromagnetic mode with an induced charge wave on the electron beam. Same approach applies also to our system with a significant difference: in the case when the DBE condition is satisfied, all four degenerate EM modes rather than one interact synchronously with the induced charge wave. In similar manner, we distinguish our approach from overmoded SWS designs [34,35] thanks to the degeneracy condition.

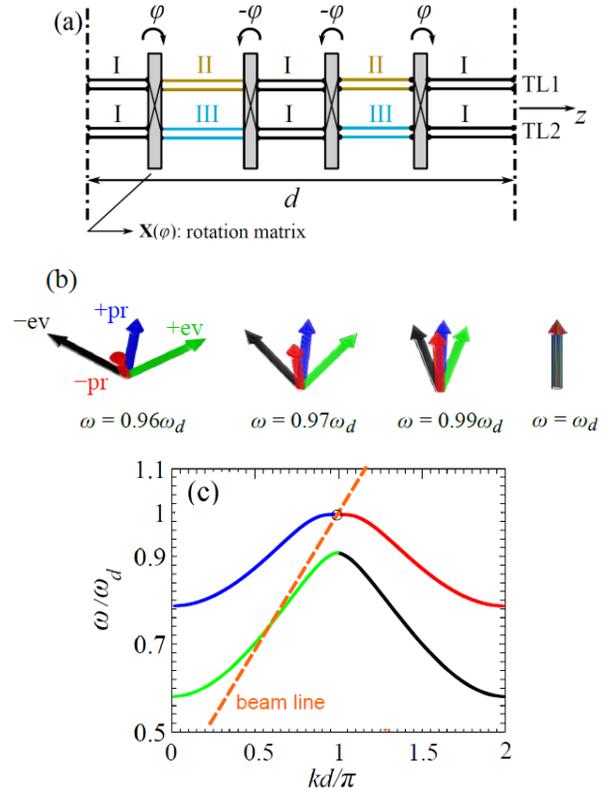

FIG. 2. (a) Equivalent representation of a unit cell of the waveguide in Fig. 1 in terms of 2-TL with a rotation matrix **X**. The CTL is designed to mimic the dispersion behavior in Fig. 1(b). (b) Arrows depicts a pictorial representation of the four, four-dimensional eigenvectors $\begin{bmatrix} V_1 & V_2 & I_1 & I_2 \end{bmatrix}^T$ relative to the four modes of the cold periodic CTL showing their evolutions from being independent to degenerate by changing frequency near the DBE. Here, $\pm$pr and $\pm$ev represent two propagating and two evanescent modes, respectively, and $\pm$ denotes the $\pm z$-direction (forward and backward waves, respectively). (c) Dispersion diagram of the cold periodic 2-TL that develops a DBE at the same frequency as the periodic waveguide in Fig. 1 for $\varphi = \varphi_{DBE} = 68.8°$. The beam line is designed to intersect with the DBE of the "cold" structure to achieve the super synchronous condition.

### A. "Cold" (with no electron beam) waveguide with a DBE

In Fig. 1(a) we propose an example of a circular, all metallic waveguide periodically loaded with metallic elliptical rings that supports a DBE. The rings have orientation misalignment between their major axes in the *x-y* plane by an angle $\varphi$. For the parameters of the waveguide reported in [26] (see also Appendix A), we show the Brillouin zone dispersion relation $\text{Re}(k)-\omega$ of four modes (two in each $\pm z$-direction, i.e., forward and backward) in the "cold" periodic structure in Fig.

1(b) calculated using a full wave simulation based on the finite elements method implemented by CST Microwave Studio. Moreover, we show symmetric positive and negative branches of Re(k). The dispersion relation of the four modes near $\omega_d$ is asymptotically equivalent to $(\omega_d - \omega) \cong h(k - k_d)^4$, where $h$ is a constant that depends on the geometry (considering the parameters in Appendix A one obtains $h = 1.76 \times 10^4$ m$^4$s$^{-1}$ when fitting the asymptotic dispersion relation to the propagating modes' dispersion obtained from full-wave simulation). In Fig. 1, we show only the branches of the dispersion diagram that have vanishing Im(k) as was done in [21,24,26]. The waveguide is designed to operate in the S-band, and the DBE frequency is 1.741 GHz.

The lower frequency modes exhibit an RBE, which is the standard band edge condition in periodic structures, whereas the higher order modes exhibit a DBE when the rings misalignment angle is precisely designed, i.e., when $\varphi = \varphi_{DBE} = 68.8°$ in this case. The reason that such waveguide in Fig. 1(a) develops a DBE in its dispersion diagram is due to a sufficient coupling between four modes (two in each $\pm z$-direction). It is achieved when two polarizations are sufficiently mixed throughout the unit cells, as explained in detail in [26]. It follows that when the rings in Fig. 1 are physically misaligned by $\varphi = 90°$ the unit cell has a 90 degrees rotational symmetric in the *x-y* plane, and we obtain two degenerated modes with orthogonal polarizations [26]. However, DBE can be attained by systematically tuning the misalignment angle $\varphi$ while all the other structural parameters are fixed (see more discussion in [26]). As a result, we identify the range of the misalignment angles where DBE can be manifested. For example, when $\varphi$ is close to 90 or 0 degrees, we obtain only an RBE due to insufficient mode mixing. We calculate the derivatives of the dispersion relation $\partial^n \omega / \partial k^n$ at $k = k_d$ for $n$ = 1, 2, 3 and 4 for different $\varphi$ values. Subsequently, we identify the precise angle $\varphi = \varphi_{DBE}$ at which the first three derivatives vanish while the fourth derivative is non-zero $\partial^4 \omega / \partial k^4 \neq 0$. This designates the condition under which DBE dispersion relation $(\omega_d - \omega) \approx h(k - k_d)^4$ is fulfilled, and it is found according to the parameters in Appendix A to be precisely at $\varphi = \varphi_{DBE} = 68.8°$ up to three significant digits. It is important to point out that if the geometry changes or the design parameters are scaled, the DBE condition will not occur at the same angle as here, but rather the precise value of misalignment angle $\varphi_{DBE}$ can be found using a similar methodology. The resulting DBE mode has a strong axial electric field component that exhibits enhanced interaction with the electron beam. In the following, we show the potential of this structure to enhance the gain for a weak electron beam. To this aim, we first elaborate on the dispersion diagram of an equivalent CTL that supports modes with the same dispersion of those in the circular waveguide exhibiting a DBE.

### B. Model for a "cold" CTL with DBE

Note that in each waveguide section with an elliptic ring, the characteristic modes are the same except for an angular rotation. We recall the basic formalism of the equivalence between transverse electric and magnetic fields to voltages and currents in a transmission line that guides the propagation along the z-direction [20,36]. Here, we are interested in four fundamental modes that are associated with a degeneracy condition, thereby we represent the periodic SWS as a cascade of *coupled* TLs, whose voltages and currents are defined with vectors $\mathbf{V}(z) = [V_1(z) \quad V_2(z)]^T$ and $\mathbf{I}(z) = [I_1(z) \quad I_2(z)]^T$, where $T$ denotes transpose, and the subscripts 1 and 2 refer to quantities in TL1 and TL2 in Fig. 2. The evolution of TL voltages and currents are related to each other by the CTL telegrapher equations [37]

$$\partial_z \mathbf{V}(z) = -\left(\underline{\underline{\mathbf{R}}} + j\omega\underline{\underline{\mathbf{L}}} - j\frac{1}{\omega}\underline{\underline{\mathbf{C}}}_c^{-1}\right)\mathbf{I}(z) \quad (1)$$
$$\partial_z \mathbf{I}(z) = -j\omega\underline{\underline{\mathbf{C}}}\mathbf{V}(z).$$

where $\partial_z \equiv \partial/\partial_z$ and $\underline{\underline{\mathbf{L}}}, \underline{\underline{\mathbf{C}}}, \underline{\underline{\mathbf{C}}}_c$, and $\underline{\underline{\mathbf{R}}}$, are $2 \times 2$ matrices representing the per-unit length inductance, capacitance, cutoff capacitance and resistance of the CTL segment. Note that the cutoff capacitance matrix $\underline{\underline{\mathbf{C}}}_c$ is included in the formulation to account for cutoff condition in the waveguide at low frequencies, [20] mimicking the behavior of TM$^z$ (transverse magnetic to z) modes in a circular waveguide [38]. Indeed, such high pass characteristic is observed in the full wave simulation of the loaded circular waveguide in Fig. 1. To describe the evolution of voltages and currents in the 2-TL system in Fig. 2, from one CTL segment to another, an interface transfer matrix that couples the fields on these two adjacent CTLs needs to be considered. In general, CTLs can have internal coupling, i.e., the per-unit length CTL parameters $\underline{\underline{\mathbf{L}}}, \underline{\underline{\mathbf{C}}}, \underline{\underline{\mathbf{C}}}_c$, and $\underline{\underline{\mathbf{R}}}$, have off-diagonal elements representing inductive or capacitive coupling between TLs. However for the sake of simplicity, we consider now an example of an CTL where TLs are uncoupled in each segment, i.e., $\underline{\underline{\mathbf{L}}}, \underline{\underline{\mathbf{C}}}, \underline{\underline{\mathbf{C}}}_c$, and $\underline{\underline{\mathbf{R}}}$, are diagonal and the only coupling is provided by the interface between two adjacent segments as shown in Fig. 2(a). We also assume that reactive coupling produced by evanescent fields that may be excited at interface discontinuities in Fig. 1 are approximately accounted for in the distributed TL parameters, since the TL segments are electrically short. Accordingly, the interface coupling matrix is given by

$$\begin{bmatrix} \mathbf{V}(z_+) \\ \mathbf{I}(z_+) \end{bmatrix} = \begin{pmatrix} \underline{\underline{\mathbf{Q}}}(\varphi) & 0 \\ 0 & \underline{\underline{\mathbf{Q}}}(\varphi) \end{pmatrix} \begin{bmatrix} \mathbf{V}(z_-) \\ \mathbf{I}(z_-) \end{bmatrix} \quad (2)$$

where $\underline{\underline{\mathbf{Q}}}(\varphi)$ is a rotation matrix that mixes the TL voltages and currents across the interface between two contiguous CTL segments, $z_-$ and $z_+$ denote the longitudinal coordinates just before and after the CTL interface, respectively, and $\varphi$ is the mixing (misalignment) parameter. When the field's phase change is ignored across an interface we may write $\underline{\underline{\mathbf{Q}}}(\varphi)$ as

$$\underline{\underline{\mathbf{Q}}}(\varphi) = \begin{pmatrix} \cos(\varphi) & -\sin(\varphi) \\ \sin(\varphi) & \cos(\varphi) \end{pmatrix}. \quad (3)$$

The coupling between TL1 and TL2 (at interfaces or within CTL segments) is necessary to establish a DBE. Bloch modes in Fig. 2(b) and (c) in a periodic CTL whose unit cell is shown in Fig. 2(a) are found by resorting to a transfer matrix formulation, as described in the next section and in Appendix B. We point out that $\varphi$ here emulates the effect of the misalignment angle of the rings axes in the actual waveguide in Fig. 1, thus allowing for the DBE tuning procedure described above.

## C. Model for a CTL interacting with an electron beam

In order to develop the interaction model of an SWS as the one in Fig. 1, which has a DBE, with an electron beam we adopt an CTL approach that extends the single mode interaction theory developed by Pierce and contemporaries [4,14,39,40] to multimode interaction as was done in [20,33,41]. We assume that the electron beam has a very small cross section, and that it is infinite along the $z$-direction, neglecting any transverse motion of electrons and fringing effects due to the structure's finite length. That allows to treat the electron beam as a smoothed-out flow of charges [14,33]. All other assumptions are also stated in [20]. For consistency we focus on the CTL in Fig. 2 (a) whose unit cell consists of five segments of 2-TL (Appendix A), that are connected in such a way that mode mixing guarantees the presence of a DBE.

The interaction of SWS modes with an electron beam is taken into account by investigating the induced charge waves as in [14,33]. Such charge wave describes the bunching and debunching of electrons, which causes energy exchange between the beam and the modes in the SWS. The electron beam has an average (d.c.) current $I_0$ and a d.c. equivalent beam kinetic voltage $V_0$ that is related to the average electron velocity $u_0$ by $V_0 = u_0^2/(2\eta)$, and $\eta = e/m = 1.758 \times 10^{11}$ C/kg is the electron charge-to-mass ratio, where $-e$ and $m$ refer to the charge and mass of the electron, respectively. The electromagnetic fields in the waveguide induce a perturbation (modulation or disturbance) on the electron beam [14] described by a modulation of the charge wave current $I_b$ and modulation of beam velocity $u_b$ with an equivalent kinetic voltage modulation $V_b = u_0 u_b/\eta$ [20,33], with the same frequency as the electromagnetic fields in the SWS. Therefore the total beam current is $I_0 + I_b$ with $|I_b| \ll I_0$ and the total equivalent beam voltage is $V_0 + V_b$ with $|V_b| \ll V_0$, based on small-signal considerations. We introduce **s** as two dimensional vector coupling coefficient between the beam current modulation $I_b$ and an equivalent displacement current injection into the CTL, causing amplification of waves in the CTL. Similarly, we introduce **a** as two dimensional vector coupling coefficient between the voltage acting as a force and the charge wave, causing bunching of the charge wave. The model can also conveniently adapt to the realistic properties of the waveguide such as DBE mode distribution (by considering the transverse field eigenmodes in each waveguide segment [20]).

We define the plasma frequency, $\omega_p$, as $\omega_p^2 = n_v e\eta/\varepsilon_0 = 2V_0 u_0/(\varepsilon_0 A)$ with $n_v$ being the volumetric electron density, and $A$ is the beam area. When the beam area is finite (corresponding to non-vanishing plasma frequency) the self-generated forces within the electron beam causes debunching of electrons or the space-charge effect.

The evolution along the $z$ direction of the time-harmonic electromagnetic wave (described in terms of TL voltages and currents) as well as the charge wave current and voltage modulations, $I_b$ and $V_b$, respectively, is described using our CTL approach [42] based on the following coupled CTL-beam evolution equations,

$$\partial_z \mathbf{V}(z) = -\left[j\omega \underline{\underline{\mathbf{L}}}(z) + \underline{\underline{\mathbf{R}}}(z) - j\frac{1}{\omega}\underline{\underline{\mathbf{C}}}^{-1}_{=c}(z)\right]\mathbf{I}(z)$$

$$\partial_z \mathbf{I}(z) = -j\omega \underline{\underline{\mathbf{C}}}(z)\mathbf{V}(z) - \partial_z[\mathbf{s}(z)I_b(z)]$$

$$(j\omega + u_0 \partial_z)V_b(z) = -u_0 \partial_z[\mathbf{a}^T(z)\mathbf{V}(z)] + j\frac{2V_0 \omega_p^2}{I_0 \omega}I_b(z)$$

$$(j\omega + u_0 \partial_z)I_b(z) = j\omega \frac{I_0}{2V_0}V_b(z), \quad (4)$$

where the last two equations above describe the charge wave dynamics coupled to the CTL through the coupling coefficients **a** and **s**. For simplicity, in the following we choose $\mathbf{a}^T = \mathbf{s}^T = [1\ 1]$ which means that both TLs interact with the charge wave. We conveniently define a space varying *state vector* composed only of the field quantities that vary along the $z$-direction, which are the transmission line voltage and current vectors, as well as the charge wave current and voltage modulations, $I_b$ and $V_b$, respectively:

$$\Psi(z) = \begin{bmatrix} \mathbf{V}^T(z) & \mathbf{I}^T(z) & V_b(z) & I_b(z) \end{bmatrix}^T.$$ The linear set of equations in (4) are then conveniently rewritten as a first order partial differential transport equation

$$\partial_z \Psi(z) = -j\underline{\underline{\mathbf{M}}}(z)\Psi(z), \quad (5)$$

where $\underline{\underline{\mathbf{M}}}(z)$ is the 6×6 system matrix that describes all the $z$-dependent CTL, electron beam and space-charge parameters, including coupling effects and losses [33]. (More details are in Appendix A.) Within each segment of the unit cell in Fig. 2 we assume that $\underline{\underline{\mathbf{M}}}(z)$ is $z$-invariant, corresponding to a homogenous cross section within each segment, i.e., the CTL is described by constant parameters (inductances, capacitances and losses per unit length) as well as the coupling parameters **a** and **s**. Therefore, within each segment $\underline{\underline{\mathbf{M}}}$ is independent of $z$, and the solution of (5) is in the form $\Psi(z) = \exp[-j(z - z_0)\underline{\mathbf{M}}]\Psi(z_0)$ with $\Psi(z_0)$ being a boundary condition at coordinate $z_0$ in the same segment or at its boundary. The 6×6 transfer matrix of each CTL segment is then constructed as $\underline{\mathbf{T}}(z, z_0) = \exp(-j(z - z_0)\underline{\mathbf{M}})$. Accordingly the relation of the state vector at the boundaries $z_n$ and $z_{n+1}$ of each segment is given by

$$\Psi(z_{n+1}) = \underline{\mathbf{T}}(z_{n+1}, z_n)\Psi(z_n), \qquad (6)$$

where here we assume that $z_{n+1} > z_n$. An interface between two homogenous CTL segments (interfaces between different waveguide cross-sections or discontinuities in the waveguide) is described by a 6×6 coupling matrix

$$\underline{\mathbf{X}}(\varphi) = \begin{pmatrix} \underline{\mathbf{Q}}(\varphi) & 0 & 0 \\ 0 & \underline{\mathbf{Q}}(\varphi) & 0 \\ 0 & 0 & \underline{\mathbf{1}} \end{pmatrix}, \qquad (7)$$

which is written in terms of a 2×2 rotation block matrix $\underline{\mathbf{Q}}(\varphi)$ that mixes the fields (i.e., voltages and currents) across the interface between two contiguous CTL segments. In principle, $\underline{\mathbf{X}}(\varphi)$ can account for all the modal interactions at the discontinuity. Here we simply neglect the higher order evanescent modes that may be excited due to this discontinuity thus representing electromagnetic fields using only the two propagation/evanescent modes in the 2-TL in Fig. 2. A more quantitatively precise analysis should employ a sufficient number of modes to make sure evanescent modes are well accounted for, where we can represent $\underline{\mathbf{X}}(\varphi)$ by reactive lumped components. We assume that the charge wave is continuous across the interface between two segments that is represented by the 2×2 identity matrix block $\underline{\mathbf{1}}$.

Since the periodic CTL comprises homogenous segments (Fig. 2), we have a matrix $\underline{\mathbf{M}}_m$ for each of the five segment CTL unit cell, with $m = \{1, 2, 3, 4, 5\}$. Now we recall the transfer matrix of a segment of the CTL $\underline{\mathbf{T}}_m$ is given by $\underline{\mathbf{T}}_m = \exp(-j\underline{\mathbf{M}}_m d_m)$ and we consider a periodic CTL as in Fig. 2(a). We then obtain the transfer matrix of the unit cell by cascading the transfer matrices of individual segments using the group properties of the transfer matrix [24]. For example, for the CTL under study in Fig. 2, we write the total transfer matrix of the unit cell as

$$\underline{\mathbf{T}}_U = \underline{\mathbf{T}}_1 \underline{\mathbf{X}}(\varphi) \underline{\mathbf{T}}_2 \underline{\mathbf{X}}(-\varphi) \underline{\mathbf{T}}_1 \underline{\mathbf{X}}(-\varphi) \underline{\mathbf{T}}_2 \underline{\mathbf{X}}(\varphi) \underline{\mathbf{T}}_1 \ . \qquad (8)$$

The transfer matrix approach for a 6-dimensional state vector leads to a 6×6 transfer matrix for the unit cell, $\underline{\mathbf{T}}_U$. The state vector evolves across the unit cell as

$$\Psi(z + d) = \underline{\mathbf{T}}_U \Psi(z), \qquad (9)$$

where $d$ is the period (see Figs. 1 and 2). This transfer matrix approach is used to obtain both the Bloch modes for an infinitely long cold CTL and to calculate the transfer function for CTL with finite length interacting with the electron beam. The main focus in the following is to explore unconventional giant gain scheme associated with a finite length DBE structure (such as the structure in Fig. 1 or analogous ones [26]) interacting with an electron beam, whereas more details of the modes in an infinitely long interactive system are found in [20].

### D. Dispersion for a "cold" CTL

For the cold CTL modeling, we set the charge wave current $I_b = 0$ in (4) and thus consider only the first 4×4 block of the unit cell transfer matrix described in the previous subsection. Accordingly, the eigenvalues of transfer matrix $\underline{\mathbf{T}}_U$ are related to the Bloch wavenumbers of the modes in the "cold" periodic structure (Appendix B, Eq. (B2) and (B3)). Fig. 2(b) depicts a graphical sketch of the four eigenvectors (two propagating, in the $\pm z$-directions i.e., forward and backward, and two evanescent modes, decaying in the $\pm z$-directions) schematically represented in a simpler three dimensional space. Here, pr and ev stand for propagating and evanescent modes, respectively, and + and − signs denote forward and backward modes, respectively. These modes are independent below and above the DBE frequency, and they coalesce at $\omega = \omega_d$ as depicted, causing a fourth order degeneracy. For the CTL parameters provided in Fig. 2 and Appendix A, the corresponding dispersion diagram of the four Bloch modes supported by this cold CTL is shown in Fig. 2(c). The DBE frequency occurs at $\omega_d = 2\pi(1.741) \times 10^9 \text{ s}^{-1}$. The dispersion diagram is obtained assuming $\varphi = \varphi_{\text{DBE}} = 68.8°$ in the rotation matrix $\underline{\mathbf{X}}(\varphi)$ in (7). Other TL parameters are chosen such that the dispersion diagram of the unit cell of 2-TL mimics that of the circular waveguide in Fig. 1(b) obtained from full wave simulations, in terms of cut off frequency, band edge frequency, as well as DBE condition. At frequencies less than the DBE frequency, the periodic CTL structure supports two propagating modes whose Bloch wavenumbers are real and denoted by $k_{\pm \text{pr}}$ and two evanescent modes with complex wavenumbers $k_{\pm \text{ev}}$. Evanescent modes are not shown in Fig. 2(c) because they require complex wavenumber representation, as shown in [20]. These eigenmodes coalesce at the DBE and therefore independent solutions for the state vectors can be only found using a generalized bases composed of diverging non-Bloch modes (see Appendix C and more discussion in Ref. [24]).

In Fig. 3 we show the dispersion diagram when the $\varphi$ in (3) is slightly varied from $\varphi_{\text{DBE}} = 68.8°$ using the equivalent CTL model. These results show how the DBE condition is perturbed, i.e., we show how the curves deviate from the exact DBE condition in Fig. 2(c). Varying the misalignment parameter $\varphi$ from 66 to 72 degrees, i.e., detuning from the DBE by ~ ±4 %, results in curves that are still close to those of the DBE condition, i.e., still very flat near $k_d = \pi/d$ though the ideal DBE condition is no longer satisfied. These deviations can occur when considering tolerances or faults in a realistic implementation of the waveguide, and considered within the limit of small perturbation. The effect of detuning is apparent where the flatness of dispersion is deteriorating near the DBE for misalignment parameter different than $\varphi_{\text{DBE}}$, however the deviation is still small. For comparison, we also show the cold structure dispersion diagram obtained using full-wave simulation (Fig. 3 inset) very close to the DBE condition. We report good agreement with the CTL simulations in the trend of such detuning effect on the dispersion diagram, near the DBE. All simulations in the rest of this paper are carried out using the CTL method in Secs. II-B (hot structure) and II-C (cold structure).

## III. GIANT GAIN VIA SUPER SYNCHRONIZATION

### A. Four mode super synchronization

The new amplification regime proposed in this paper is based on the four degenerate modes in a SWS at the DBE (super synchronism condition). The beam line in Figs. 2-3 defined by $k = \omega/u_0$ represents the dispersion relation associated to a charge wave induced on the electron beam, whose electrons travel at an average velocity of $u_0$. The periodic 2-TL system as that in Fig. 2 is able to support a DBE, i.e., the degeneracy between four Bloch modes. Near the band edge ($k \approx k_d = \pi/d$, and $\omega \approx \omega_d$), there are four *k*-solutions of the equation $[k - k_d]^4 \cong [(\omega_d - \omega)/h]$: two modes are propagating and two modes are evanescent, whose wavenumbers are $k_{\pm\text{pr}}(\omega) \cong k_d \mp [(\omega_d - \omega)/h]^{1/4}$ and $k_{\pm\text{ev}}(\omega) \cong k_d \mp j[(\omega_d - \omega)/h]^{1/4}$, respectively, for frequencies lower than the DBE frequency. (Here the fourth order root is defined as real and positive.) Note that each of those Bloch modes possesses an infinite number of spatial Floquet harmonics (slow and fast) of which only a slow one is synchronized with the electron beam with an average velocity of $u_0$. The wavenumbers of these four synchronous spatial harmonics are equal to $k_d = \pi/d$, and are degenerate at the band edge angular frequency $\omega_d$, as shown in Figs. 1-3. Therefore, the super synchronism condition is satisfied when

$$u_0 \approx \frac{\omega_d}{k_d} = \frac{\omega_d d}{\pi}. \qquad (10)$$

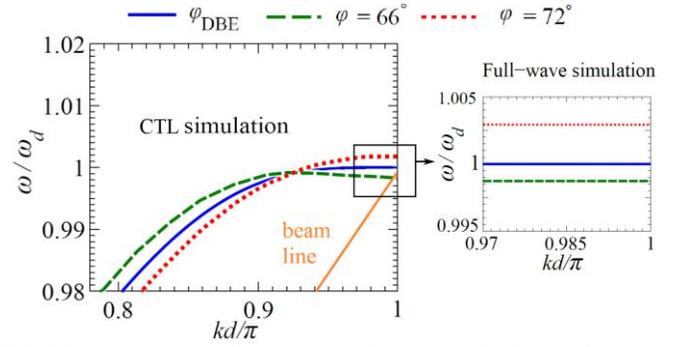

FIG. 3. Important branch of the dispersion diagram near DBE radian frequency $\omega_d$ for the cold equivalent CTL structure in Fig. 2(a), for $\varphi = 66°$ and $72°$ as well as $\varphi_{\text{DBE}}$. The inset shows a zoomed plot of the dispersion diagram obtained using full-wave simulation for the same detuning of the angle $\varphi$.

This condition is represented in Fig. 2(c) by the intersection of the "cold" structure dispersion and the beam line, at the band edge. This is the pre-requisite for the interaction scheme proposed in this paper based on the four degenerate modes synchronization, which is achieved either by varying the beam average velocity $u_0$ or by changing the SWS to exhibit the DBE condition at a designed frequency. In proximity of the DBE frequency four Floquet harmonics, one for each of these four degenerate modes, constructively engage in the interaction with the electron beam because they all have approximately the same phase velocity $\omega_d/k_d$. When structure perturbations affect the SWS, the DBE condition is not perfectly encountered also at the DBE frequency. In this case there are almost identical, yet slightly distinct four Bloch eigenmodes, therefore the interaction process of electron beam and every mode is still guaranteed in case a small structural perturbations.

Furthermore, the dispersion diagram of the hot structure is slightly perturbed when the average electron beam current is not high [20]. On the other hand, it is completely distorted for large beam currents and synchronism may not explicitly be casted into the form of (10) implying that one may lose the degeneracy and the benefits of the DBE features for amplification. Here we will consider only weak beam currents since this is the case when the advantage is anticipated. In summary, equation (10) is the main criterion for the design of a DBE electron beam amplifier that can be finely tuned using the dispersion diagram of the hot structure and resonance condition due to a finite length of the SWS. Fine tuning of the electron beam velocity allows for optimizing the gain. Indeed, super synchronization in the finite length would require a slightly different value of beam velocity $u_0$ compared to the one obtained from (10) as we show in the next subsection.

### B. Giant gain in finite length structures with DBE

We consider now a finite length structure composed of *N* unit cells of the CTL in Fig. 2(a), without electron beam (i.e., in a cold periodic SWS). The four port network representation of the CTL circuit is shown in Fig. 4 with the termination impedance as well as a voltage generator. We also assume the charge wave (the small signal modulation of the electron beam) enters in the interaction area (at $z = 0$ in Fig. 4) with zero pre-modulation, i.e., $I_b(0) = 0$ and $V_b(0) = 0$, that are used as boundary conditions for equations (4), i.e., for $\Psi(z = 0)$ in (5). The other boundary conditions for $\mathbf{V}(z = 0)$ and $\mathbf{I}(z = 0)$ depend on the generator and its internal impedance at $z = 0$ (Fig. 4). A general setting of boundary conditions includes different source/load topologies than the single source and load example in Fig. 4. Choosing different excitation and load impedances will not affect the conclusions drawn in the paper; also manipulating such boundaries may even be beneficial for optimizing the structure's gain. Moreover, imposing the charge wave boundary conditions $I_b$ and $V_b$ at $z = 0$ indicates that such boundary conditions provide a propagating charge wave excited by the EM wave that does not encounter reflection from the "collector" side in Fig. 4. Accordingly, a unique state vector $\Psi(z)$ at any $z$ is evaluated by conventional transfer matrix method described in the previous section and in Appendix B, as well as in Refs [20,33]. Referring to Fig. 4 we define the TL circuit voltage transfer function as the ratio

$$T_F = V_L(\omega)/v_g, \qquad (11)$$

where $V_L(\omega)$ is the load circuit voltage at the termination impedance $Z_L$ due to the generator voltage $v_g$.

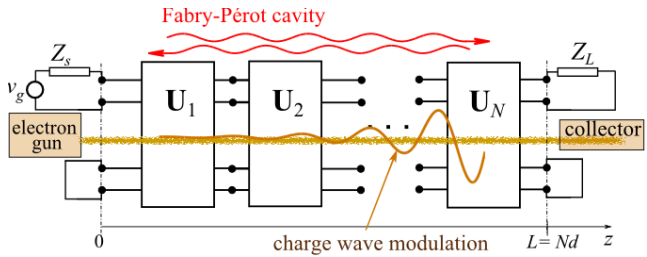

FIG. 4. Conceptual example of CTL interacting with electron beam. Sequence of N cascaded unit cells of 2-TL in Fig. 2(a) forms a Fabry-Pérot cavity. Resonance occurs when a mode travels a whole round trip with a total phase shift that is multiple of $2\pi$.

We are interested in the condition where the excited fields create a standing wave because of the Fabry-Pérot resonance whereby constructive interference of two propagating modes with opposite group velocities leads to a strong transmission resonance. Here we are investigating the Fabry-Pérot resonance closest to the DBE frequency. This resonance frequency is approximated by $\omega_{r,d} \approx \omega_d - h\left[\pi/(Nd)\right]^4$ [24,26] where the fitting constant $h$ was defined in Sec. II-A. The alignment of the Fabry-Pérot resonance with the DBE condition leads to a huge mismatch of Bloch waves at the CTL ends. Because of the degeneracy condition, this leads to a giant enhancement in the energy stored, compared to the energy escaping at the two ends of the structure, and as a consequence it leads to a huge enhancement in the Q-factor [22]. Various general characteristics of DBE relative to gain enhancement were reported in [22]. The amplification scheme discussed here is also referred to as convective instability, where amplification occurs below certain threshold current, as typical in many high power devices [32,43,44]. Note that in a *finite-length* SWS with DBE interacting with an electron beam, the maximum gain will not occur exactly at $\omega = \omega_d$, but rather at a frequency close to the transmission resonance angular frequency $\omega_{r,d}$ of the cold structure. At $\omega_{r,d}$, the cold SWS supports two propagating modes with Bloch wavenumbers $k_{+\mathrm{pr}}(\omega_{r,d}) \approx k_d - \pi/(Nd)$, for the forward mode, and $k_{-\mathrm{pr}}(\omega_{r,d}) \approx k_d + \pi/(Nd)$ for the backward mode, as well as two evanescent modes with Bloch wavenumbers $k_{\pm\mathrm{ev}}(\omega_{r,d}) \approx k_d \mp j\pi/(Nd)$ [22,24]. Therefore we choose the beam velocity $u_0$ is so that it match the phase velocity of the forward propagating mode such that $u_0 \cong \omega_{r,d}/k_{+\mathrm{pr}} = \omega_{r,d}/(k_d - \pi/(Nd))$, and this coincides with (10) for large $N$. Such value of $u_0$ has been selected in our specific design of the finite length TWT to achieve a giant amplification, without seeking a thorough optimization. In other words, fine tuning of $u_0$ around $\omega_{r,d}/k_d$ is necessary to maximize the gain but not attempted here for simplicity. Nonetheless, the cold SWS is still undergoing a four-mode degeneracy at $\omega_{r,d}$; in the sense that the excited SWS develops giant growth of the amplitude of field that is constructed from the four degenerate modes (refer to Fig. 6). Therefore the concept of four mode synchronization is still valid in the finite length structure, even when we chose slightly different $u_0$ than what would condition (10) provides, thanks strong excitation of all four modes at resonance.

Indeed, for large N, we use $\omega_{r,d} \cong \omega_d$ therefore we will always assume a constant $u_0$ for the sake of simplicity in all cases. In the following we take $u_0 = 0.497c$ where $\omega_d/k_d \cong 0.47c$.

In Fig. 5(a) we show the small-signal voltage gain $= 20\log|T_F|$ in dB when the electron beam has average current $I_0 = 20$mA in the frequency range near the DBE where the highest enhancement in gain occurs. For comparison purposes, we show the gain in dB for two different cavity lengths, i.e., for N equal to 21 and 25 unit cells, and observe how the gain varies with misalignment parameter $\varphi$ around the DBE design. The peak gain is slightly different and occurs at slightly different frequencies, all very close to $\omega_d$, for the three misalignment angles.

We also show that the gain is substantially enhanced when the number of unit cells increases from 11 to 25 for the same average beam velocity $u_0 = 0.497c$. To better quantify this feature, we report in Fig. 5(b) the gain in dB varying as a function of the number of unit cells N, for the three misalignment parameter (the gain for each case, varying as function of length and misalignment parameter, is calculated at the frequency where the peak occurs in Fig. 5(a)). The gain in dB increases monotonically for increasing $N$ and the largest gain (in the range of $N$ shown in Fig. 5(b)) is observed for the optimum misalignment parameter $\varphi_{\text{DBE}}$. Moreover, the enhancement in gain follows the dramatic decrease in the group velocity. It can be shown that the gain enhancement is directly proportional to the increase in the group delay [22]. Following this proposition, we superimposed a fitting formula

$$\text{Gain}_{\text{DBE}} \sim -b + aN^4 \text{ [dB]}, \quad (12)$$

to the gain calculated via transfer matrix for optimum misalignment parameter $\varphi_{\text{DBE}}$ in Fig. 5(b). Here $a$ and $b$ are fitting constants proportional to the beam current, while $b$ also accounts for the initial losses due to the boundary conditions imposed on the TL at $z = 0$, similar to the conventional Pierce model for very small space-charge effect [4,14]. In our case we obtain the fitting constants as $a \cong 5.1 \times 10^{-5}$ and $b \cong 4.9$. The R-squared value, which is a number between 0 and 1 and a statistical measure [45] of how well equation (12) represents the calculated gain for optimum misalignment parameter $\varphi_{\text{DBE}}$, is found to be ~0.995 demonstrating an accurate fit since it is very close to 1. Note that for small values of $N$ ($N < 15$ in Fig. 5(b)) there is no gain since the length of the interactive system is not sufficient to cause significant bunching that leads to amplification for an average beam current of 20 mA. Moreover, in the DBE amplifier, four synchronous degenerate modes are engaged with the electron beam and the amplification is a result of such four mode interactions. Therefore the circuit voltage gain is calculated for all the modes excited in the structure, since the generator would excite all growing modes due to super synchronism. However, in the Pierce type TWT, only one exponentially growing (amplifying) mode exists so naturally the gain will be dominated by that mode. The voltage gain, according to Pierce formula [4,14], accounts only for that sole growing mode, which results in the fact that $\text{Gain}_{\text{Pierce}} \propto N$ [14,39]. In DBE amplifiers, since we have four degenerate modes, $\text{Gain}_{\text{DBE}} \propto N^4$. On one hand, we point out once again that this regime corresponds to a convective instability operation of the amplifier (i.e., the amplifier is stable in the zero-drive mode with no input of electromagnetic signal in the CTL). On the other hand, the amplifier stability with large signal drive is another important aspect that should be addressed in the future, by accounting for nonlinearities and other practical imperfections. Note that increasing the TWT's length, with end reflections, beyond a critical length can cause the amplifier to enter an absolute instability regime and oscillate (a typical scenario in mismatched TWTs [29,30,46]). Therefore, DBE amplifiers can be readily designed when careful choices of beam current, length as well as loads are considered, whereas oscillators can also be conceived using different topologies of the DBE cavities.

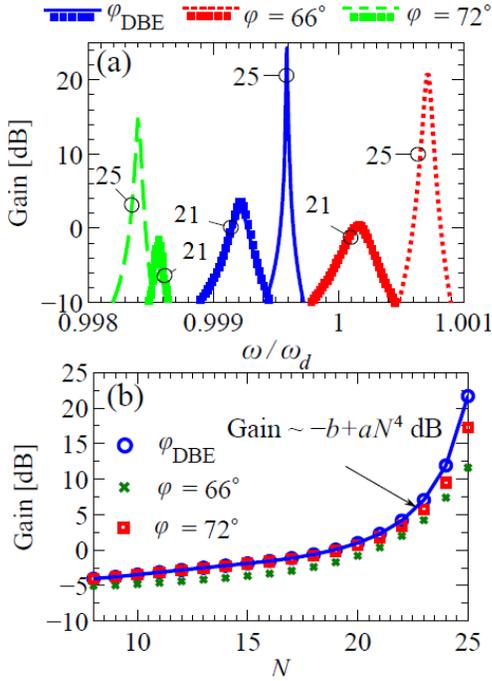

FIG. 5. (a) Gain, $20\log|T_F|$, of the CTL system in Fig. 4, assuming $I_0 = 20$ mA and $u_0 = 0.497c$ near the DBE frequency for different misalignment angles and number of unit cells indicated by numbers in the figure. (b) Gain scaling versus the number of unit cells $N$ for three misalignment angles, showing an accurate fitting of gain to $\text{Gain}_{\text{DBE}} = -b + aN^4$ [dB].

To further elaborate the phenomena of the giant amplification indicated in (12) and Fig. 5, we show in Fig. 6 the mode distribution in the FPC, when the electromagnetic mode interacts with an electron beam with the average beam current of $I_0 = 3$ [mA] ($V_0 = 60.1$ KV). We show the total voltage $\|\mathbf{V}(z)\| = \left(|V_1|^2 + |V_2|^2\right)^{1/2}$ and current $\|\mathbf{I}(z)\| = \left(|I_1|^2 + |I_2|^2\right)^{1/2}$, as well as the magnitude of the beam modulation (the charge wave) voltage $V_b$ and current $I_b$, along the FPC length for a FPC made of $N = 32$ unit cells. The profiles shown in Fig. 6 are calculated by sampling those aforementioned fields at the beginning of each unit cell, and we also assume that those profiles are excited using a unit voltage generator as in Fig. 4. The TL and charge wave voltages and currents are shown for three cases, for the exact DBE design with $\varphi = \varphi_{\text{DBE}} = 68.8°$ and the two other misalignment parameter. Each case is plotted at an angular frequency where the maximum gain occurs (Fig. 5), close to $\omega_d$.

It is important to note the strong beam modulation (charge wave) of both voltage $V_b$ and current $I_b$ in each of the three designs and especially for the case with $\varphi_{\text{DBE}}$ (blue curve). The higher gain obtained with the DBE design in Fig. 5, with $\varphi = \varphi_{\text{DBE}} = 68.8°$, is mainly attributed to a more pronounced standing wave in Fig. 6 where the voltages, current and beam modulation are stronger than those with other misalignment parameter. Note that the beam modulation profiles $V_b$ and $I_b$ do not display the standing wave form as the one displayed by voltage and current in the transmission line but rather a monotonic growth; which is a consequence of the associated boundary conditions of the system in which the beam is allowed to exit at the collector side with no reflection effects. Vice versa, $\|\mathbf{V}(z)\|$ and $\|\mathbf{I}(z)\|$ show a strong standing wave behavior because of the reflection of the FPC ends. Due to the FPC configuration this gain is calculated in a narrow band corresponding to DBE transmission resonance band (the frequency band where the gain peaks in Fig. 5). However techniques to enhance the bandwidth of operation, such as chirping [47], could be explored. Furthermore, another possible utilization of the DBE could be based on exploiting the interaction between an electron beam and only backward waves.

Small signal gain of the DBE amplifiers with three different misalignment is compared in Fig. 7 with that obtained with a single and homogenous TL interacting with the beam according to the Pierce model [14,48]. The frequency of operation for each point in the plot is chosen as the frequency of the maximum peak of the transfer function in Fig. 5 which was pre-computed.

The parameters of the single and homogenous transmission line are chosen based on the average of those we use for the 2-TL system in Fig. 2, and are given in Appendix A. In this way we recover the gain by using the small signal gain formula derived by Pierce for synchronous operation (with negligible space-charge effect) [14]: $G = -9.54 + 47.3 C_p L_g$ [dB], where $C_p$ is the Pierce coupling parameters and $L_g$ is the number of wavelengths of the amplifier. Results in Fig. 7 show that the gain calculated using small beam current for the case with $\varphi_{DBE}$ is significantly larger than that obtainable with a Pierce single-TL TWT (it is 10 times larger, when the beam current is $I_0 = 3$ mA) of the same length. We should stress that large gains are usually obtained in a homogeneous TL according to the standard Pierce model [14,39,48,49], when the TWT attains a significant length. Analogously large gain is obtained in long periodic structures, working far away from edge of the band edge [6,50]. In these designs usually the interaction impedance is high, the beam current is intense, and the structure is long. However large gain with the super synchronism operation condition shown in this paper is obtained with shorter TLs and weaker beam currents as shown in Fig. 7. Practical aspects of the DBE amplifier including power extraction and preventing unwanted oscillations are important and should be studied in the future. The former can be addressed by carefully designing load and matching networks for CTLs that allow for optimum extraction of the power from the amplifier, especially for high levels of output RF power. Unwanted oscillations can be mitigated by proper placement of attenuators and severs [2], for example.

## IV. CONCLUSION

We have proposed a new superior amplification regime in slow-wave structures occurring under the degenerate band edge (DBE) condition for the system's dispersion relation. This regime is based upon on super synchronous operation of the electron beam and the four degenerate EM modes near the DBE. The regime exhibits superior gain enhancement that cannot be achieved in conventional TW amplifier devices that are based on a single EM synchronous operation with the electron beam. As an example we have proposed a circular loaded waveguide that develops a DBE in its dispersion diagram and developed its 2-TL model to understand and explore the modal characteristics and the EM modes interaction with charge waves. We have shown a giant gain scaling with length that is promising for enhancement of microwave power amplifiers performance and efficiency. The super-synchronization proposed here has also potential application in reduction of starting oscillation beam current as compared to conventional oscillators. This can be beneficial for enhanced efficiency of high power microwave sources.

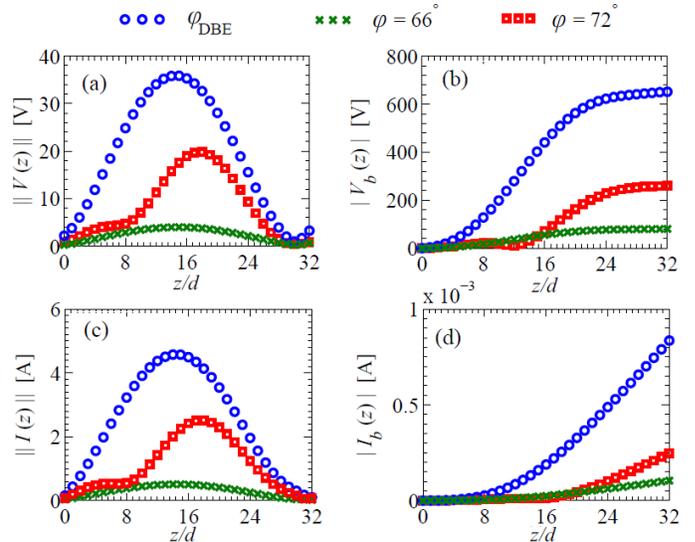

FIG. 6. Magnitudes of (a) total voltage, (b) total current, (c) beam voltage modulation, and (c) beam current modulation, for the amplified resonant mode near $\omega_d$ for different misalignment parameter with $I_0 = 3$ [mA].

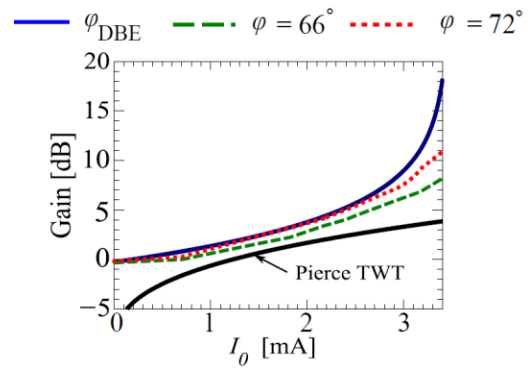

FIG. 7. Small signal gain $20\log|T(\omega)|$ for the finite length SWS coupled to the electron beam in Fig. 3, versus the electron beam current. Gain is calculated at the resonance frequencies of the FPCs in Fig. 4, with different misalignment angles. Results show much larger gain than that obtained with the Pierce model for a single TL SWS with the same length $Nd$, and $N = 32$.

## ACKNOWLEDGMENT

This research was supported by AFOSR MURI Grant FA9550-12-1-0489 administered through the University of New

Mexico. Authors acknowledge also support from AFOSR Grant FA9550-15-1-0280. They would like to thank CST Inc. for providing CST Microwave Studio that was instrumental in this work.

## APPENDIX A: PARAMETERS USED IN NUMERICAL SIMULATIONS

The parameters of the waveguide in Fig. 1 are as follows: ring length 15 mm, rings separation 2.5 mm, ring thickness 5 mm, major radius 25 mm, and aspect ratio 2.5:1. The periodicity along the *z*-direction is $d = 40$ mm. The equivalent CTL parameters for TL I are $L = 0.33$ μH/m, $C = 2.9$ nF/m, $C_c = 80$ fF.m, and $d_I = 2.5$ mm (the middle segments length is $d_I/2$), whereas for TL II are $L = 3.1$ μH/m $C = 0.3$ nF/m, $C_c = 0.2$ pF/m and $d_{II} = 15$ mm, and for TL III L are 0.6 nH/m, $C = 0.3$ nH/m $C_c = 0.2$ pF/m and $d_{III} = 15$ mm. Those parameters are extracted from the full-wave simulation by fitting the dispersion relation in Fig. 2(c) to that of the waveguide in Fig. 1, and maintaining an average impedance of the unit cell constituent TLs of ~51.6 Ω. For data in Fig. 5 only, we have assumed $\omega_p = 0.02\omega_d$ and a distributed loss in all TLs of $10 \text{m}\Omega/\text{m}$. The comparison with a Pierce type TWT is done with computing the gain in a transmission line whose characteristic or interaction impedance is equal to the average of the impedance of all the TL in the CTL structure (51.6 Ω), with the same beam parameters.

## APPENDIX B: ON THE ADOPTED CTL-BEAM INTERACTIVE SYSTEM FORMULATION

In our CTL model we have ignored relativistic effects on the electron beam, and if we wish to include those effects in our computation, it will lead to less than 4% correction in velocity calculation. The relativistic factor $\gamma = \left[1-\left(u_0/c\right)^2\right]^{-1/2}$ is about ~1.13 whereas we assumed no relativistic effects for simplicity, i.e., $\gamma = 1$. The beam voltage $V_0$ is defined as $V_0 = u_0^2/(2\eta)$ [2] and we have used $V_0 = 60.1$ kV. Such value was typically used in many experimental studies, as those in [44,51,52] that may result in beam currents up to few amperes. The matrix $\underline{\mathbf{M}}_m$ in (5) for the *m*th CTL segment is given by

$$\underline{\mathbf{M}}_m = \begin{bmatrix} \underline{\mathbf{0}} & -j\underline{\underline{\mathbf{R}}}_m + \omega\underline{\underline{\mathbf{L}}}_m - \mathbf{C}_{cm}^{-1}/\omega & \mathbf{0} & \mathbf{0} \\ \omega\underline{\underline{\mathbf{C}}}_m & \underline{\mathbf{0}} & \omega\eta\frac{\rho_0}{u_0^2}\mathbf{s}_m & -\beta_0\mathbf{s}_m \\ 0 & \mathbf{a}_m^T\left(-j\underline{\underline{\mathbf{R}}}_m + \omega\underline{\underline{\mathbf{L}}}_m - \mathbf{C}_{cm}^{-1}/\omega\right) & \beta_0 & -\frac{\omega_p^2}{\omega\rho_0\eta} \\ 0 & 0 & -\omega\eta\frac{\rho_0}{u_0^2} & \beta_0 \end{bmatrix} \quad (B1)$$

Considering an infinitely long periodic CTL system as in Fig.2, we seek periodic solutions of the state vector $\mathbf{\Psi}(z)$ in the Bloch form $\mathbf{\Psi}(z+d) = e^{-jkd}\mathbf{\Psi}(z)$, where *k* is the complex Bloch wavenumber. Then we write the eigensystem as,

$$\underline{\mathbf{T}}_U \mathbf{\Psi}_n = e^{-jk_n d}\mathbf{\Psi}_n, \quad (B2)$$

where $\mathbf{\Psi}_n$ is the *n*th eigenvector whose corresponding eigenvalue is $e^{-jk_n d}$ with $n = 1,2,...,6$. The eigenvalues are evaluated by solving the characteristic equation

$$\det\left[\underline{\mathbf{T}}_U - e^{-jkd}\underline{\mathbf{1}}\right] = 0, \quad (B3)$$

for complex *k*, which has 6 possible complex valued *k* at each real frequency. More involved details of the $k-\omega$ dispersion diagram of the coupled system (4) are found in [20]. Here equation (B3) is used to generate Fig. 2(c) when we solve the decoupled system, i.e., when the electron beam does not interact with the SWS therefore finding the dispersion of the "cold structure" as well as the dispersion of the naked beam (that is calculated numerically by setting $\mathbf{s}=\mathbf{a}=0$). However, when the beam is interacting with the SWS of finite length, we always assume that $\mathbf{s}(z) = \mathbf{a}(z) = \begin{bmatrix} 1 & 1 \end{bmatrix}^T$ (beam interacts with both the TLs) and use the model in Sec. II.

## APPENDIX C: FOURTH ORDER DEGENERACY AND GENERALIZED EIGENVECTORS AT THE DBE

A DBE condition represents a fourth order degeneracy of the eigenvectors of the system. In other equivalent terms, at DBE the cold structure transfer matrix's eigenvalue has an algebraic multiplicity of four and a geometric multiplicity of one [21,22,24,53]. Here we would like to emphasize the special features of the eigenvectors at the DBE to show that it is a remarkable degeneracy condition. Consider the eigenvalue problem in (B2), yet for the cold structure (without beam interaction) hence it is written in term of a four dimensional state-vector $\mathbf{\Psi}_n = \begin{bmatrix} \mathbf{V}^T(z) & \mathbf{I}^T(z) \end{bmatrix}^T$, and the $4 \times 4$ unit cell transfer matrix of the cold structure $\underline{\mathbf{T}}_u$ as

$$\left[\underline{\mathbf{T}}_u - e^{-jk_n d}\underline{\mathbf{1}}\right]\mathbf{\Psi}_n = 0, \quad (C1)$$

where $\mathbf{\Psi}_n$ are the regular eigenvectors, found at the reference plane *z* = 0 and $\underline{\mathbf{1}}$ is the 4×4 identity matrix (see Appendix C in [20]). At the DBE, the four Bloch wavenumbers are $k_n = k_d = \pi/d$, with n=1,2,3,4, such that $e^{-jk_n d} = -1$, and $\underline{\mathbf{T}}_u$ is similar to a 4×4 Jordan block [54]. Because of the Jordan block similarity, (C1) would provide *only one* independent periodic solution for eigenvectors $\mathbf{\Psi}_n$ in the system with a fourth order degeneracy. Accordingly, the system's four independent state vector solutions at the DBE must be found using generalized eigenvectors bases as [54,55]

$$\left[\underline{\mathbf{T}}_u + \underline{\mathbf{1}}\right]^n \mathbf{\Psi}_n = 0, \quad n = 1,2,3,4. \quad (C2)$$

We find one regular eigenvector with index *n* = 1 and three generalized eigenvectors with indices *n* = 2, 3 and 4.

Mathematical details of generalized eigenvectors can be found in a number of textbooks, for example, see Refs. [55,56,54]. All eigenvectors correspond to an eigenvalue of $e^{-jk_nd}=-1$. Clearly this indicates that these four degenerate modes (constructed by generalized eigenvectors) form one Bloch (periodic) mode $\Psi_1$ and three non-periodic modes diverging along the +z-direction as $z$, $z^2$, and $z^3$ respectively, as dictated by degeneracy [24]. Such diverging eigenmode characteristics are essential for understanding the giant resonance effects in cavities with modal degeneracies [24,53]. When operating at the angular resonance frequency $\omega_{r,d}$ defined in Sec. III-B (that is close to $\omega_d$) the modes of the structure are no longer degenerate however they are constructed from a perturbed version of those diverging modes (exactly degenerate) with a non-vanishing group velocity [53]. In other words, when the SWS is excited at $z = 0$ by the voltage generator in Fig. 4 at the resonance $\omega_{r,d}$, close to the DBE frequency, such diverging modes are not strictly speaking excited, but being close to that condition generates a growing behavior of the voltage along the z-axis seen in Fig. 6 up until the cavity center. Under this quasi-DBE condition, the boundary condition at $z = 0$ (i.e., loads and voltage generators) enforces eigenvectors associated to propagating and evanescent modes to be excited with giant weight, as discussed in [22,24].